\documentstyle[aps,epsfig,prl,multicol]{revtex}
\begin{document}
\draft
\title{Manifestation of the Arnol'd Diffusion in Quantum Systems}
 \author{V.Ya. Demikhovskii$^1$, F.M. Izrailev$^2$ and A.I. Malyshev$^1$}
 \address{$^1$ Nizhny Novgorod State University, Gagarin Ave., 23, 603600,
Nizhny Novgorod, Russia}
 \address{$^2$ Instituto de F\'isica, Universidad Aut\'onoma de Puebla,
 Apdo. Postal J-48, 72570, Puebla, Mexico} \maketitle

\begin{abstract}
We study an analog of the classical Arnol'd diffusion in a quantum
system of two coupled non-linear oscillators one of which is
governed by an external periodic force with two frequencies. In
the classical model this very weak diffusion happens in a narrow
stochastic layer along the coupling resonance, and leads to an
increase of total energy  of the system. We show that the quantum
dynamics of wave packets mimics, up to some extent, global
properties of the classical Arnol'd diffusion. This specific
diffusion represents a new type of quantum dynamics, and may be
observed, for example, in 2D semiconductor structures (quantum
billiards) perturbed by time-periodic external fields.
\end{abstract}


\begin{multicols}{2}
\narrowtext

As is known, the main mechanism for the onset of dynamical chaos
in classical Hamiltonian systems is the destruction of
separatricies of non-linear resonances due to perturbation terms.
However, for a small enough perturbation in conservative systems
with two degrees of freedom ($N=2$), as well as for
one-dimensional time-dependent Hamiltonians, chaotic regions are
bounded in the phase space by the Kolmogorov-Arnol'd-Moser (KAM)
surfaces \cite{LL92}. In this case the chaos is non-global since
chaotic trajectories are located in restricted regions of the
phase space of a system.

The situation changes drastically for many-dimensional systems
($N>2$), where KAM surfaces no longer separate one stochastic
region from another, and chaotic layers generically form a {\it
stochastic web} \cite{comm}. Thus, if trajectory starts in the
vicinity of a specific non-linear resonance, it can diffuse along
stochastic layers of many resonances that cover the whole phase
space. This universal global instability known as the {\it Arnol'd
diffusion} \cite{Arnold}, for the first time was observed in
numerical experiments \cite{CKS71}, and then studied in detail in
\cite{Izr,old} (see also review \cite{C79} and the book
\cite{LL92}).

It is very difficult to observe this specific diffusion since its
rate is exponentially small, and it manifests itself only for
initial conditions inside very narrow stochastic layers. However,
it is believed that the Arnol'd diffusion may play an important
role in different physical situations. For example, it is argued
that this diffusion is crucial for the estimate of the stability
of our Solar system, and is responsible for a loss of electrons in
magnetic traps (see in \cite{C79}). Also, the Arnol'd diffusion
may have a strong influence for the dynamics of protons in high
energy storage rings. Recently, the possibility of observation of
the Arnol'd web for a Hydrogen atom in crossed electric and
magnetic fields has been discussed in Ref.\cite{MDU96}.

So far, all the studies of the Arnol'd diffusion refer to
classical models. On the other hand, it is important to understand
what is the fingerprint, if any, of this diffusion in quantum
systems. This question is far from trivial since quantum effects
are known to typically suppress classical effects of an
exponentially weak diffusion\cite{shuryak}. In this Letter we
perform a detail study of the quantum Arnol'd diffusion in a
simple model, and show what are peculiarities of quantum dynamics
that are due to this phenomenon.

The system under consideration consists of two coupled quartic
oscillators one of which is perturbed by a two-harmonic force.
Following to Refs.\cite{Izr} where the corresponding classical
model was studied both analytically and numerically, we write the
Hamiltonian,
\begin{equation}
\label{ham}
\hat H=\hat H^0_x+\hat H^0_y-\mu xy-f_0 x(\cos\Omega_1t+
\cos\Omega_2 t).
\end{equation}
Here
\begin{equation}
\label{hatH^0_x}
\hat H^0_x=\frac{\hat p_x^2}2+\frac{x^4}4, \qquad \hat
H^0_y=\frac{\hat p_y^2}2+\frac{y^4}4,
\end{equation}
and $\mu, f_0, \Omega_1, \Omega_2$ stand for the coupling
constant, amplitude and two frequencies of an external force,
respectively. For momentum and coordinate operators the standard
commutation relations are assumed, $[\hat p_x,\,x]=-i\hbar_0$,
$[\hat p_y,\,y]=-i\hbar_0$, with $\hbar_0$ as the dimensionless
Plank constant.

In the classical model the separatrix of the main coupling
resonance $\omega_x^0 \approx \omega_y^0$ (determined in the
absence of perturbation, $f_0=0$) is destroyed and creates a very
narrow stochastic layer. The Arnol'd diffusion is caused by two
driving terms with commensurate frequencies $\Omega_1$ and
$\Omega_2$, that ``force" a chaotic trajectory diffuse {\it along}
the resonance layer \cite{Izr,C79}.

To consider the quantum model, first, one should find quantum
eigenstates corresponding to this coupling resonance, for $f_0=0$.
Then, by switching on the perturbation, the analysis can be
performed by using the Floquet formalism, since the perturbation
is periodic in time. In this way, one can construct the evolution
operator that allows one to study the dynamics of the model.

In order to find stationary states corresponding to the coupling
resonance, we write the wave function in the form,
\begin{equation}
\label{psi_x_y}
\psi(x,\,y)=\sum_{n,m} c_{n,m}\psi^0_n(x)\psi^0_m(y).
\end{equation}
Here $\psi^0_n(x)$ and $\psi^0_m(y)$ are the eigenfunctions of
$\hat H^0_x$ and $\hat H^0_y$ (calculated numerically), and
coefficients $c_{n,m}$ satisfy to the stationary Schr\"odinger
equation,
\begin{equation}
\label{stat eq}
Ec_{n,m}=(E_{n}+E_{m})c_{n,m}-
\mu\sum_{n',m'} x_{n,n'}y_{m,m'}c_{n',m'},
\end{equation}
where $E_{n}$ and $E_{m}$ are the eigenvalues of the Hamiltonians
$\hat H^0_x$ and $\hat H^0_y$.

Our interest is in the dynamics of the model in the vicinity of
the main coupling resonance determined by the condition
$\omega_{n_0}=\omega_{m_0}$. Here $\hbar_0\omega_{n_0}=E_{n_0}'$,
$\hbar_0\omega_{m_0}=E_{m_0}'$, and $n_0=m_0$ defines the
resonance centre. In this region one can expand $E_n$ and $E_m$ in
the Tailor series, keeping second order terms. It is convenient to
introduce the indexes $p=k+l$ and $k$, where $k=n-n_0$ and
$l=m-m_0$. Then, one can write,
\begin{equation}
\label{system}
\matrix{
Ec_{k,p}=\left[\hbar_0\omega p+E_{n_0}''\left(k^2-pk+{p^2\over 2}
\right)\right]c_{k,p}-\cr
-\mu\Bigl(\dots + \sum_{k'} x_{k,k'}y_{p-k,-1-k'}\,c_{k',-1}+
\Bigr. \cr
\quad\;\;\,+ \sum_{k'} x_{k,k'}y_{p-k,-k'}\,c_{k',0}+ \cr
\Bigl. \qquad\qquad\;\:\,+ \sum_{k'} x_{k,k'}y_{p-k,1-k'}\,c_{k',1}+
\dots \Bigr),
}
\end{equation}
with $\omega \equiv \omega_{n_0}$. Since matrix elements $x_{m,n}$
and $y_{m,n}$ of coordinates are equal to zero for transitions
between states of the same parity, the solution of Eqs.
(\ref{system}) consists of two independent sets, for odd and even
values of $p$.

\begin{figure}[tb]
\begin{center}
\mbox{\psfig{file=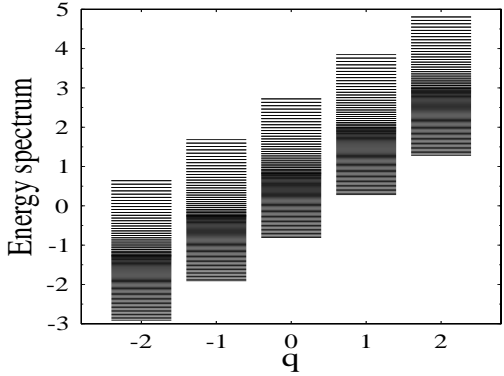,width=8cm,height=6cm}}
\end{center}
\caption{Energy spectrum of the system (\ref{system})
in units of $\hbar_0\omega$. Five groups with $121$ states in each
group are shown for $\mu=10^{-4}$, $\hbar_0\approx 1.77\cdot
10^{-5}$, and $n_0=446$.}
\label{spectrum}
\end{figure}

Let us now consider the case of a small nonlinearity, when the
condition $\hbar_0\omega p\gg E_{n_0}''\left(k^2-kp+{p^2\over
2}\right)$ is satisfied. The corresponding numerical data for a
fragment of the energy spectrum is shown in Fig.\ref{spectrum}.
One can see that the spectrum consists of series of energy levels,
that are shifted one from another by the value $\hbar_0\omega$.
The structure of energy spectrum in each group is typical for a
quantum nonlinear resonance \cite{BVI87}. Lowest levels are
practically equidistant with the spacing equal to $\hbar_0\tilde
\omega$, where $\tilde \omega$ is the classical frequency of small
phase oscillations at coupling resonance. Accumulation points
correspond to classical separatricies, and all energy levels
inside separatricies are non-degenerate. The states slightly above
separatricies are quasi-degenerate due to the symmetry of a
rotation in opposite directions.

In accordance with the spectrum structure it is convenient to
characterize the states at coupling resonance by two indexes, by
the group number $q$ and by the index $s$ which stands for the
levels inside groups. Then, the energy in each group can be
written as $E_{q,s}=\hbar_0
\omega q+E^M_{q,s}$, where $E^M_{q,s}$ is the Mathieu-like spectrum
of one group.

Now we analyze the dynamics of the model in the presence of the
external force. For commensurate frequencies $\Omega_1$ and
$\Omega_2$, the perturbation is periodic with the period $T=i
T_1=j T_2$, where $T_1=2\pi/\Omega_1$, $T_2=2\pi/\Omega_2$ and
$i$, $j$ are integers. The initial conditions were taken for a
system to be about half-way between the two driving resonances,
$\omega = (\Omega_1+\Omega_2)/2$.

Since the Hamiltonian (\ref{ham}) is periodic in time, the
solution of the Schr\"odinger equation can be written as
\begin{equation}
\label{Flok}
\psi(x,y,t)=\exp\left( -{i\varepsilon_Qt\over \hbar_0}\right)
u_Q(x,y,t).
\end{equation}
Here $u_Q(x,y,t)=u_Q(x,y,t+T)$ and $\varepsilon_Q$ are the
quasienergy (QE) functions and quasienergies, respectively. They
are determined by the evolution operator $\hat U$ describing the
dynamics of our system in one period of the external field,
\begin{equation}
\label{ev_1}
\hat U u_Q(x,y)=\exp\left( -{i\varepsilon_QT\over \hbar_0}\right)
u_Q(x,y).
\end{equation}
Here the argument $t$ is omitted since we are interested in the
wave function only for discrete times $NT$, with $N=1,2,...$.

It is now naturally to represent the QE functions in the form
$u_Q(x,y)=\sum_{q,s} A^Q_{q,s}\psi_{q,s}(x,y)$, where
$\psi_{q,s}(x,y)$ are eigenstates of the unperturbed Hamiltonian
$\hat H^0=\hat H^0_x+\hat H^0_y-\mu xy$. The coefficients
$A^Q_{q,s}$ are the eigenvectors of the operator $\hat U$ in the
representation of two coupled nonlinear oscillators, that can be
found by diagonalization of the corresponding matrix
$U_{q,s;q',s'}$. This matrix can be numerically obtained in the
following way. Let the evolution operator $\hat U$ act on the
initial state
$C^{(q_0,s_0)}_{q,s}(0)=\delta_{q,q_0}\delta_{s,s_0}$. Then the
wave function $C^{(q_0,s_0)}_{q,s}(T)$ at time $T$ forms the
column of the evolution operator matrix,
\begin{equation}
\label{ev_3}
\matrix{
U_{q,s;q',s'}(T)C^{(q_0,s_0)}_{q',s'}(0)=U_{q,s;q_0,s_0}(T)
=C^{(q_0,s_0)}_{q,s}(T).}
\end{equation}
Repetition of this procedure for different initial states
$C^{(q',s')}_{q,s}(0)=\delta_{q,q'}\delta_{s,s'}$ determines the
whole matrix $U_{q,s;q',s'}(T)$. As a result, the wave function
$C^{(q_0,s_0)}_{q,s}(T)$ can be computed numerically by
integration of the nonstationary Schr\"odinger equation,
\begin{equation}
\label{time eq}
\matrix{
i\hbar_0 \dot C_{q,s}=\left( \hbar_0 \omega q+E^M_{q,s}\right)
C_{q,s}- \cr -f_0\sum_{q',s'} x_{q,s;q',s'}\left(
\cos\Omega_1t+\cos\Omega_2t \right)C_{q',s'}.}
\end{equation}
If we introduce slow amplitude $b_{q,s}(t)$ via the transformation
\begin{equation}
\label{change}
C_{q,s}(t)=b_{q,s}(t)\exp\left[ -i\left( q \omega
+E^M_{q,s}/\hbar_0 \right) t\right],
\end{equation}
then after some algebra one can obtain
\begin{equation}
\label{reson}
\matrix{
i\hbar_0 \dot b_{q,s}=-f_0\cos\left({\delta \Omega \over
2}t\right) \times \cr
\times \sum_{s'}
\left[ x_{q,s;q+1,s'}\,b_{q+1,s'}\,e^{-i\bigl
(E^M_{q+1,s'}-E^M_{q,s}\bigr)t/\hbar_0 }+ \right. \cr
\left. \qquad\; +\,x_{q,s;q-1,s'}\,b_{q-1,s'}\,e^{-i\bigl
( E^M_{q-1,s'}-E^M_{q,s}\bigr)t/\hbar_0 } \right],}
\end{equation}
where $\delta \Omega = \Omega_1-\Omega_2$. Using the resonance
approximation, we keep only the most important slowly oscillating
terms with $q'=q\pm 1$.

\begin{figure}[tb]
\begin{center}
\raisebox{0.55cm}{\psfig{file=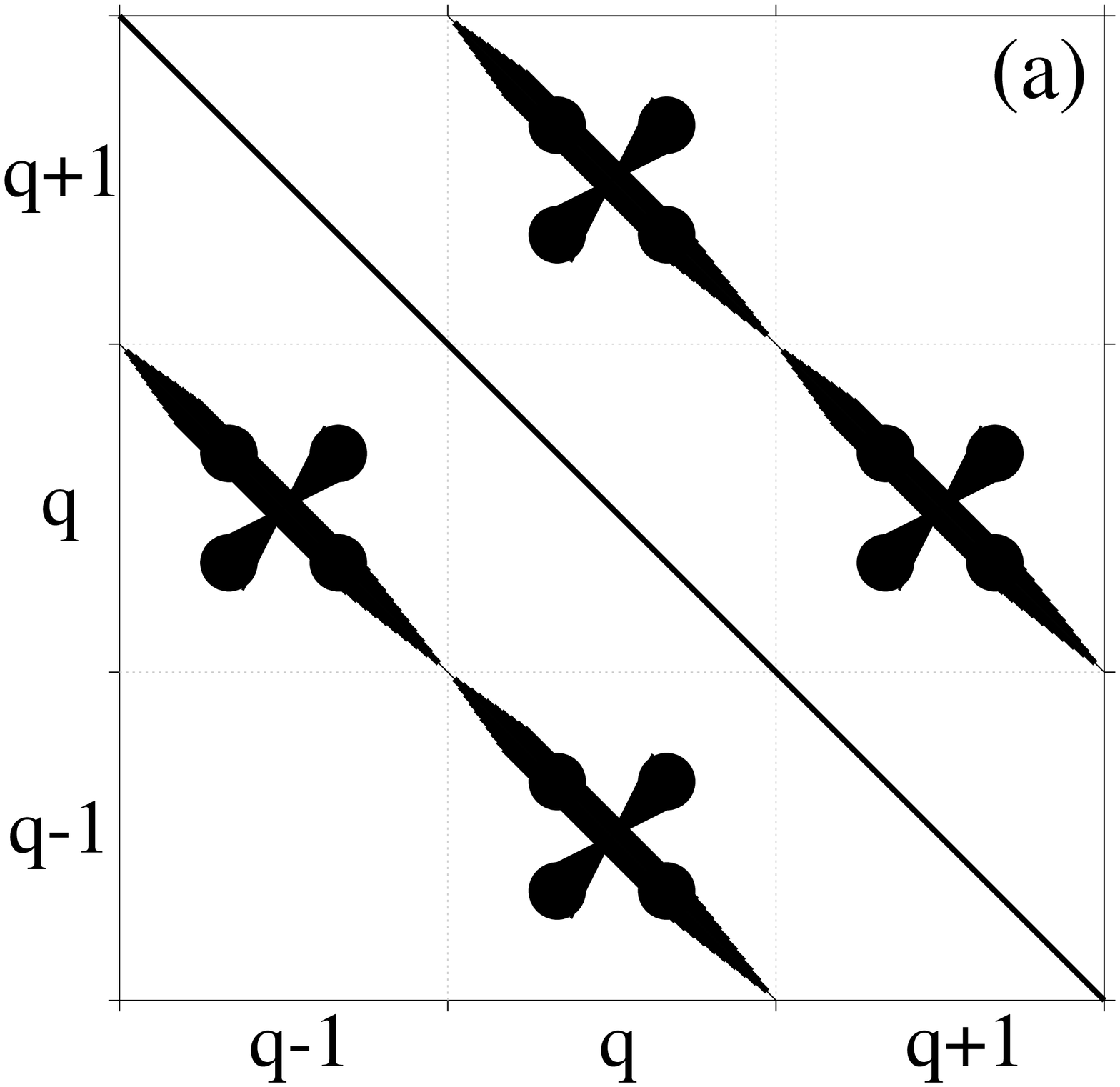,width=3.5cm,height=3.5cm}}
\mbox{\psfig{file=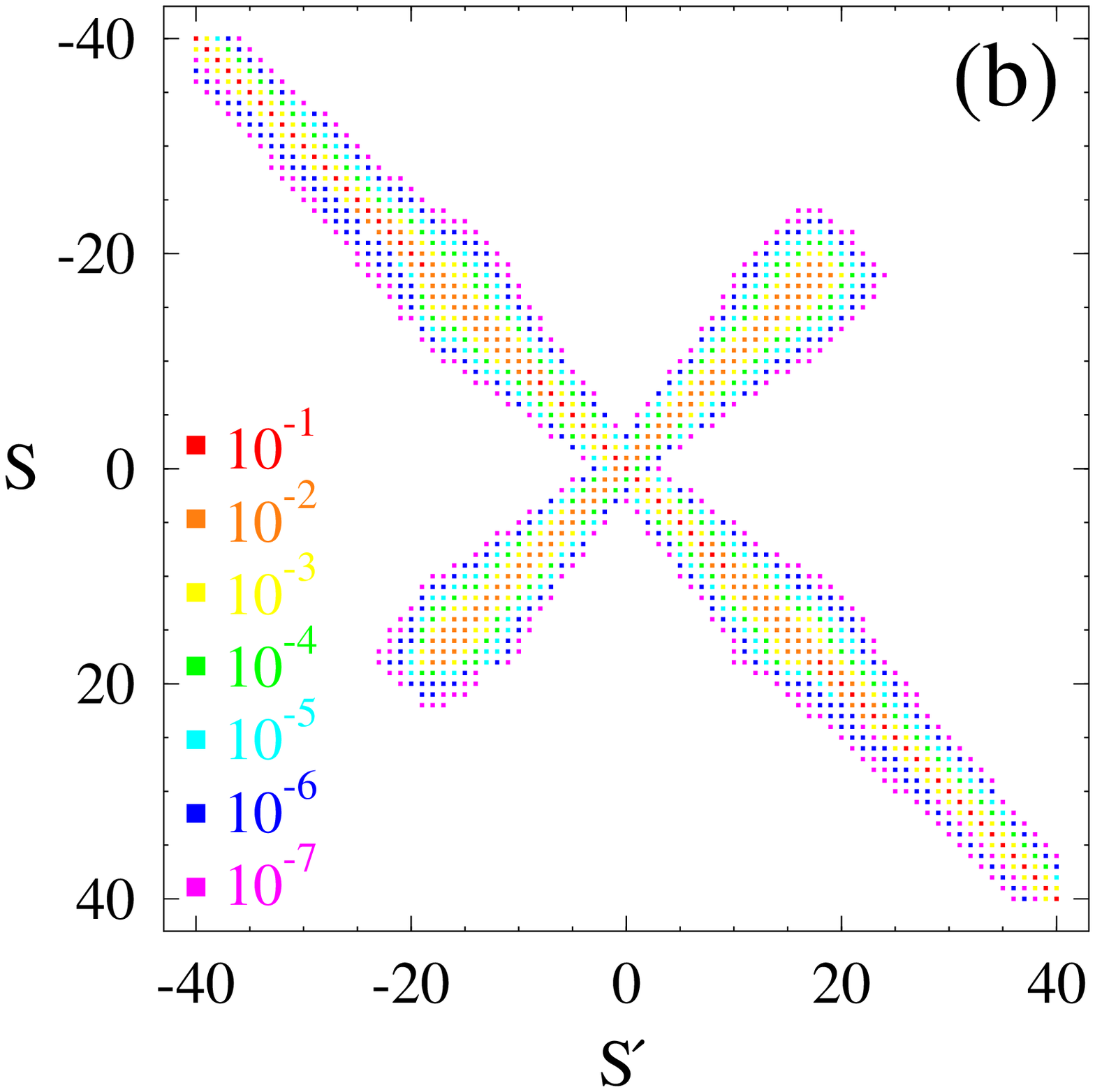,width=4.5cm,height=4.5cm}}
\end{center}
\caption{(a) Matrix elements $|x_{q,s;q\pm 1,s'}|$ that define
the transition probability along the coupling resonance, for the
same parameters as in Fig.\ref{spectrum}. (b) Detailed structure
of the \lq\lq cross\rq\rq{}-like broadening.}
\label{m_el}
\end{figure}

The matrix elements $x_{q,s;q\pm 1,s'}$ in (\ref{reson}) define
transition probabilities between the states $s$ and $s'$ from the
neighbor groups. This matrix has a specific block structure shown
in Fig.\ref{m_el}a.

Black and white regions correspond to big and small matrix
elements, respectively. One fragment of this structure is shown in
Fig.\ref{m_el}b in more details. The matrix elements in the center
of Fig.\ref{m_el}b correspond to the transitions between lowest
states (at the center of coupling resonance) of the groups with
$q=0$ and $q=1$. Matrix elements at the corners of
Fig.\ref{m_el}b, where $s,\,s'>30$ or $s,\,s'<-30$, define
transitions between the states above separatrix. All these
elements quickly decrease with an increase of the difference
$|s-s'|$. The \lq\lq cross\rq\rq{}-like broadening in
Fig.\ref{m_el}b, where matrix elements are large, corresponds to
the transitions between separatrix states. As a result, the
transition probability between separatrix states of neighbor
groups (along the coupling resonance) is much larger than the
transition probability between other states. This phenomenon is
analogous to the quantum diffusion inside a separatrix, which was
investigated in a degenerate Hamiltonian system \cite{demi}.

\begin{figure}[tb]
\begin{center}
\mbox{\psfig{file=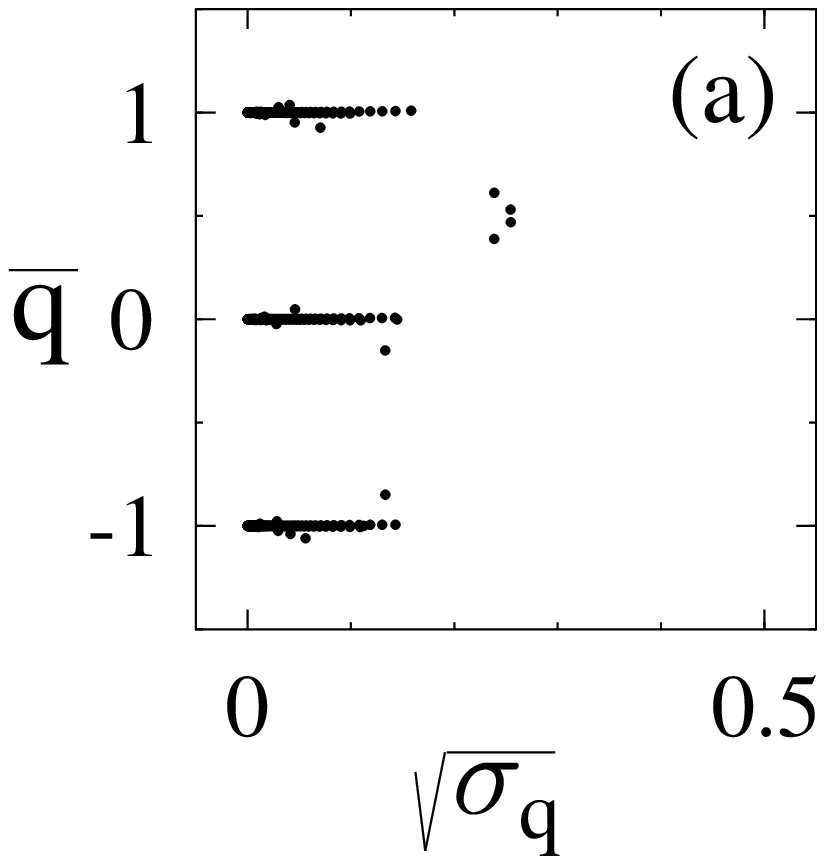,width=4cm,height=4cm}
\psfig{file=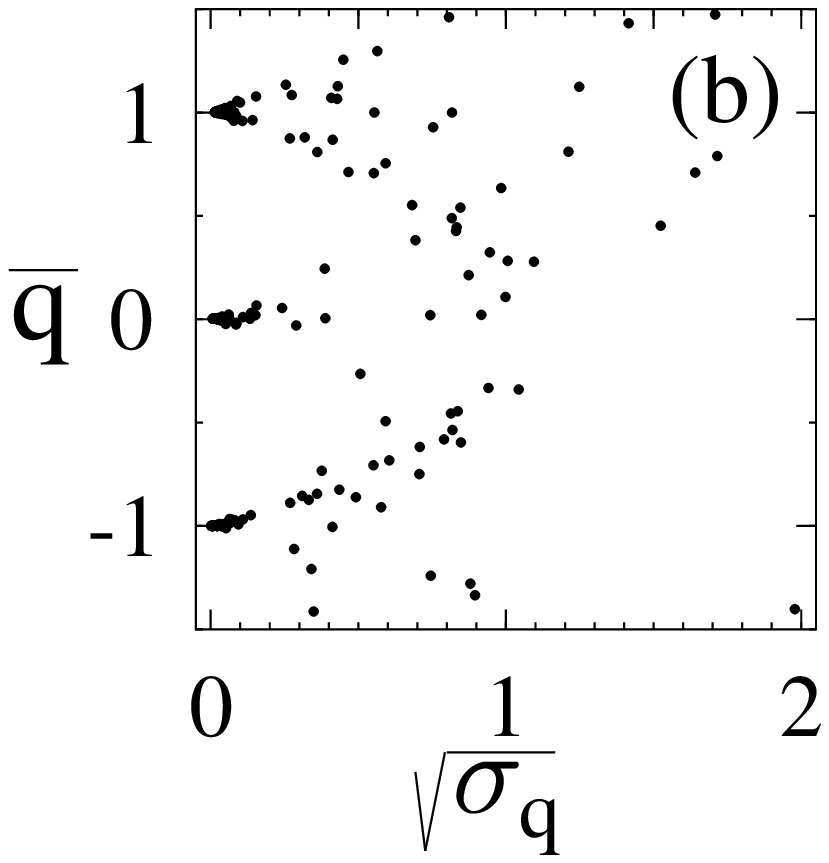,width=4cm,height=4cm}}
\end{center}
\caption{Plot of $\bar q$ versus $\sqrt{\sigma_q}$ for QE
functions $A^Q_{q,s}$ in the regions $q=0,\pm 1$ for different coupling
constants: a) $\mu=3\cdot 10^{-5}$, b) $\mu=10^{-4}$. Points correspond to
different QE functions.}
\label{cluv}
\end{figure}

Solving numerically Eqs.(\ref{reson}), we obtain the matrix
$U_{q,s;q',s'}(T)$ that determines the eigenvalues $\varepsilon_Q$
and QE functions $A^Q_{q,s}$. Global properties of QE functions
can be understood in terms of their ``centers" $\bar
q=\sum_{q}q\sum_{s}|A^Q_{q,s}|^2$, and dispersion
$\sigma_q=\sum_{q}(q-\bar q)^2\sum_{s}|A^Q_{q,s}|^2$ in the
unperturbed ($f_0=0$) basis, see Fig.\ref{cluv}. One can see that
for a relatively small $\mu$ and $f_0$, the QE functions do not
couple unperturbed states with different values of $q$. On the
other hand, a stronger coupling results in a kind of
``delocalization" that is characterized by the spread of QE
functions over many groups of states with different $q$. This fact
manifests a large probability for the transition between those
states that are involved in the diffusion along the coupling
resonance.

Direct computation of the evolution of wave packets has confirmed
the occurrence of a weak diffusion in energy space, that is
analogous to the classical Arnol'd diffusion. Numerical data have
been obtained by computing the evolution matrix according to the
relations,
\begin{equation}
\label{ev_5}
U_{q,s;q',s'}(NT)=\sum_{Q} A^Q_{q,s} {A^{Q^{\,*}}_{q',s'}}
\exp\left(
-{i\varepsilon_QNT\over \hbar_0}\right)
\end{equation}
\begin{equation}
\label{ev_6}
C_{q,s}(NT)=\sum_{q',s'}U_{q,s;q',s'}(NT)C_{q',s'}(0).
\end{equation}

We have studied quantum dynamics and calculated diffusion
coefficient for different initial states in the regime when the
values of $\mu$ and $f_0$ are small enough and $f_0/ \mu=0.01$, so
that the coupling and two driving resonances do not overlap.

\begin{figure}[tb]
\begin{center}
\hspace{-1.0cm}
\mbox{\psfig{file=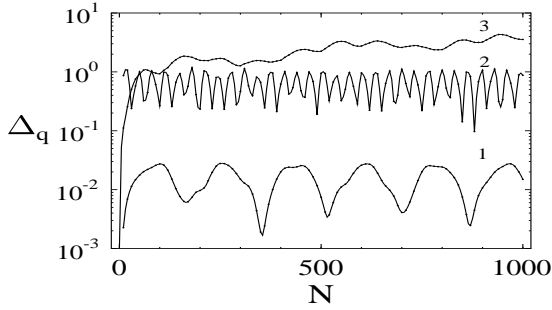,width=8cm,height=4.5cm}}
\end{center}
\caption{Time dependence of the energy dispersion for
$\mu=1.25 \cdot 10^{-4}$ and initial states: (1) near the center
of coupling resonance, (2) and (3) - above and on the separatrix.}
\label{comp}
\end{figure}

Quantum dynamics for different initial conditions is illustrated
in Fig.\ref{comp} where the energy dispersion $\Delta_q =\overline
{(\Delta H)^2}/\hbar_0^2\omega^2$ with $\Delta_q=\sum_{q}(q-
\tilde q)^2\sum_{s}|C_{q,s}|^2$ and
$\tilde q=\sum_{q}q\sum_{s}|C_{q,s}|^2$ is plotted versus the
rescaled time $N=t/T$. The figure clearly shows different
character of the evolution for three initial states. For the
states taken from below and above the separatrix, the energy width
of packets oscillates, in contrast with a diffusion-like
time-dependence for the case when the initial state corresponds to
the classical separatrix.

\begin{figure}[tb]
\begin{center}
\hspace{-1.0cm}
\mbox{\psfig{file=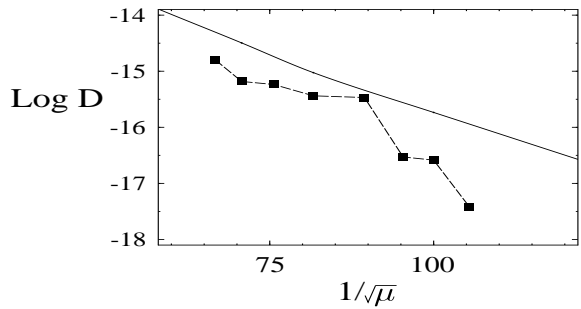,width=8cm,height=4.5cm}}
\end{center}
\caption{Quantum (squares) and classical (solid line) diffusion
coefficient vs the parameter $1/\sqrt{\mu}$. }
\label{coeff}
\end{figure}

We calculated quantum and classical diffusion coefficients and
found that the diffusion in the quantum model is systematically
weaker than the classical Arnol'd diffusion, see Fig.\ref{coeff}.
However, the global dependence of the quantum diffusion
coefficient on the control parameter $1/\sqrt{\mu}$ is similar to
the classical one. The data in this figure are obtained for
initial states corresponding to the centre of the stochastic
layer, where the diffusion coefficient is maximal. For initial
states corresponding to the border of the stochastic layer, the
diffusion coefficient strongly fluctuates and can be several times
smaller.

To count the number $M_s$ of eigenstates inside the classical
separatrix layer, we calculated the width of the classical layer
numerically, together with the number of energy levels in the
corresponding energy interval. For $\mu>1.25\cdot 10^{-4}$ the
number $M_s$ was found to be about 10. The right square in
Fig.\ref{coeff} for the lowest value of $\mu$ corresponds to the
Shuryak border
\cite{shuryak}, the latter that only one quantum state is inside
the classical stochastic layer. In this case, the classical chaos
is completely suppressed by quantum effects. We have also found
that for large times, $t>1000T$, the diffusion-like growth of the
energy terminates, thus indicating the localization of the
classical diffusion along stochastic layers.

In conclusion, we have shown that when the number $M_s$ of
stationary states inside classical chaotic separatrix layers is
relatively large, one can observe weak quantum diffusion that is
similar to the classical Arnol'd diffusion. In our model this
diffusion occurs along the nonlinear coupling resonance, due to
the influence of guiding resonances originated from the external
time-dependent perturbation.  Our results may find a confirmation
in experiments on one-electron dynamics in 2D quantum billiards
with time-periodic electric fields.

This work was supported by grants RFBR No.~01-02-17102 and
Ministry of Education of Russian Federation No.~E00-3.1-413. FMI
acknowledges the support by CONACyT (Mexico) Grant No. 34668-E.

\end{multicols}

\begin{references}
\bibitem{comm} One should distinguish the stochastic web for the
 Arnol'd diffusion from that known for systems that are linear in
 the absence of a perturbation. In the latter case the stochastic
 web arises from {\it one} resonance that have an infinite number
 of cells in the phase space, see, for example:
  A.A. Chernikov, R.Z. Sagdeev, D.A. Usikov, M.Yu. Zakharov, and
  G.M. Zaslavsky, {\it Nature}, {\bf 326}, 559 (1987).
\end{references}
\end{document}